\documentclass[12pt,preprint]{aastex}
\shorttitle{{\it Chandra} Observation of Abell 1991}

\begin{document}

% ------------------------------------------------------------

\newcommand\ergs{\rm erg\thinspace s$^{-1}\thinspace$}
\newcommand\msun{$M_{\odot}\thinspace$}
\newcommand\msunpa{$M_{\odot}\thinspace {\rm yr}^{-1}\thinspace$}

%
% ------------------------------------------------------------
%

\title{A {\it Chandra} X--ray Observation of Abell 1991: The Late 
Stages of Infall?}

\author{Mangala Sharma\altaffilmark{1},
B. R. McNamara\altaffilmark{1},
P. E. J. Nulsen\altaffilmark{2,3},
M. Owers\altaffilmark{2},
M. W. Wise\altaffilmark{4},
E. L. Blanton\altaffilmark{5,6},
C. L. Sarazin\altaffilmark{5}, 
F. N. Owen\altaffilmark{7}, and
L. P. David\altaffilmark{3}}

\altaffiltext{1}{Department of Physics \& Astronomy, Ohio University,
Athens, OH 45701;
sharma@phy.ohiou.edu, mcnamara@phy.ohiou.edu}

\altaffiltext{2}{Engineering Physics, University of Wollongong, 
Wollongong NSW 2522, Australia}

\altaffiltext{3}{Harvard-Smithsonian Center for Astrophysics, 
60 Garden St., Cambridge, MA 02138}

\altaffiltext{4}{Massachusetts Institute of Technology, 
Center for Space Research, 70 Vassar Street, Building 37, Cambridge,
MA 02139}

\altaffiltext{5}{Department of Astronomy, University of Virginia,
P. O. Box 3818, Charlottesville, VA 22903}

\altaffiltext{6}{{\it Chandra} Fellow}

\altaffiltext{7}{National Radio Astronomy Observatory, Socorro, NM 87801}

\def\mathfont#1{\ifmmode{#1}\else{$#1$}\fi} %for math font     
\def\lae{\mathrel{<\kern-1.0em\lower0.9ex\hbox{$\sim$}}}  
\def\gae{\mathrel{>\kern-1.0em\lower0.9ex\hbox{$\sim$}}}  

%
% ------------------------------------------------------------
%

\begin{abstract}

We present results from a $38$~ks {\it Chandra} X--ray observation of
the  $z=0.059$ galaxy cluster A1991. The cluster has a bright X-ray core 
and a central temperature gradient that declines inward from $2.7$~keV 
at $130$~kpc to $\approx 1.6$~keV at the cluster center.  The radiative
cooling time of the gas in the inner $10$~kpc is about $0.5$~Gyr, 
and rises to $1$~Gyr at a radius of $20$~kpc. The cooling rate of 
the gas within the latter radius is $\lae 25$~\msunpa .
The {\it Chandra} ACIS-S3 image shows that the intracluster medium has
an asymmetric surface brightness distribution with respect to the
central galaxy.  Bright knots of soft X-ray emission embedded in
a cometary structure are located approximately 10~arcsec north of the 
optical center of the cD galaxy.  
Unlike the structures seen in other cooling flow clusters,
the knots have no obvious association with the radio source.  
The structure's temperature of $0.83$~keV makes it nearly $1$~keV cooler 
than its surroundings, and its mass is $3.4 \times 10^9\ M_{\sun}$. 
Based on its bow-shaped appearance and modest overpressure with
respect to its surroundings, we interpret the structure as 
a cool mass concentration that is breaking apart as it
travels northward through the center of the cluster.

\end{abstract}

%
% ------------------------------------------------------------
%

\section{Introduction \label{sec:intro}}

The intracluster medium (ICM) in the central $\sim 100\ h^{-1}_{70}$~kpc
of many clusters of galaxies has a high density, and a radiative
cooling time that is short relative to the age of clusters.  In the
absence of heating, the hot ICM should cool, condense, and flow
inwards in a so-called cooling flow (Fabian 1994). A cooling flow
can deposit up to $\dot{M} = 1000$~\msunpa in the form of atomic and 
molecular clouds that may fuel star formation in the brightest
cluster galaxy. 
However, the observed quantities of cooler gas (e.g., Edge 2001) and 
star formation in the central galaxy (e.g., Johnstone, Fabian, \& Nulsen 1987; 
McNamara \& O'Connell 1989) are at the level of $\le 1-10\%$ of
the cooling rates determined with the {\it ROSAT} or {\it ASCA} 
X--ray missions. 
Recent high--resolution spatial and spectral data obtained with the
{\it Chandra} and {\it XMM-Newton} X--ray observatories have placed
upper limits on the cool mass deposition rates that are lower by an
order of magnitude compared to previous estimates 
(Molendi \& Pizzolato 2001; Peterson et al. 2003;
see McNamara 2002 for a review).
Though the cooling time of the gas may be as small as a few hundred
million years, very little of the ICM cools below roughly a third of
the ambient cluster temperature, 
contrary to the model prediction where the gas cools below X--ray 
temperatures. 

The cooling times and cooling rates of the ICM as well as the star
formation rates and radio powers of the central cluster galaxies span a
large parameter space. A definitive test of the cooling flow paradigm
requires comparisons of reliable, locally--determined cooling rates,
star formation rates and histories, and cold gas masses in cluster
cores spanning the range of these properties. However, the literature
contains detailed studies of mostly clusters with the largest cooling
rates and very powerful radio sources, e.g., Hydra~A (McNamara et al.~2000; 
David et al.~2001), Abell~1795 (Fabian et al. 2001), 
NGC~1275/Perseus (Fabian et al.~2000), that are extreme in many
respects. This paper discusses a galaxy cluster that hosts a moderate
cooling flow and a low--power radio galaxy, typical
of the class of cooling flow clusters as a whole. 

Abell 1991 is a galaxy cluster of Abell richness class $R=1$ at a
modest redshift of $z=0.0587$ (Struble \& Rood 1999). Its X--ray
luminosity $L_X=1.35 \times 10^{44}\ {\rm erg\ s}^{-1}$, and
temperature $T_X=5.3$~keV (White, Jones, \& Forman 1997) are
characteristic of richness $R \ge 1$ clusters.  A1991 purportedly
hosts a moderate
cooling flow of $\dot M \approx 115$~\msunpa (Stewart et al. 1984), 
and its brightest central galaxy (BCG), NGC 5778, hosts a compact, 
relatively low-power 
radio source with ${\rm log}\ P_{\rm 1400\ MHz} = 23.41\ {\rm W\ Hz^{-1}}$ 
(Owen \& Ledlow 1997). 
Only modest levels of star formation (McNamara \& O'Connell 1992)
and accreted gas ($M_{\rm HI} < 3.6 \times 10^9$\msun ; McNamara et al. 1990)
are in evidence in NGC 5778, implying that the cooling flow is largely 
in remission.

In this paper, we report the results of high spatial resolution X--ray
imaging and imaging spectroscopy of Abell 1991. 
In \S~\ref{sec:datared}, we describe the observations using the 
{\it Chandra} X--ray Observatory, and the data analysis. 
\S~\ref{sec:morph} discusses the X--ray morphology of 
the ICM.
We present the average spectral properties of A1991 in
\S~\ref{sec:avgspec}, and the spatial variation of the
X--ray temperature and abundance in \S~\ref{sec:temp_prof}.
From the spectral analysis, we
determine the distribution of the gas density, pressure and entropy in
\S~\ref{sec:nepr_prof}. We have identified a knotted substructure in 
the ICM; we discuss its nature in \S~\ref{sec:knotspec}.
We summarize and discuss our results in \S~\ref{sec:concl}. 

We employ the cosmological parameters of a flat universe with
$H_0=70$~km\ s$^{-1}$ Mpc$^{-1}$, $\Omega_m=0.3$ and
$\Omega_{\Lambda}=0.7$ throughout this paper, which gives a linear
scale of 1.14~kpc arcsec$^{-1}$ and luminosity distance of $264$~Mpc
at the redshift of A1991. 

%
% ------------------------------------------------------------
%

\section{Observation and Data Reduction \label{sec:datared}}

{\it Chandra} observations of A1991 (ObsID 3193) employed the ACIS
instrument for a total exposure time of 38~ks on 2002 December 16 and 17. 
The data and read modes were Very Faint and Timed Event,
respectively, and the focal plane temperature was $-120$~C. Five CCD
chips (ACIS-I2, I3, S1, S2 and S3) were switched on during the
observation.  In this analysis, we use data from the S3 chip only,
concentrating on the bright central $r \leq 200$~kpc region of the
cluster. 

We reduced the level~1 event data received from the {\it Chandra}
X--Ray Center pipeline. We used CIAO Version 2.3 and calibration
products CALDB Version 2.21. To convert counts to event PI values
and photon energies, we used the gain file
acisD2000-08-12gainN0003.fits in the CALDB. 
We modeled the secular gain change in ACIS-S3 using
the program and calibration data of Vikhlinin et al.\footnote{Available
at http://hea-www.harvard.edu/$\sim$alexey/acis/tgain/}
We excluded bad pixels, bad columns, and node boundaries.  
We retained only events with {\it ASCA} grades of 0, 2, 3, 4, and 6. 
During the observation, the particle background was generally stable with 
no flares, and there were no periods of bad aspect. Since the central region 
of the cluster is bright, we opted for a less conservative strategy in 
cleaning the lightcurve and retained data from the entire length of the
observation. For observations performed in Very Faint mode, 
a filter\footnote{See 
http://cxc.harvard.edu/cal/Acis/Cal\_prods/vfbkgrnd/index.html}
proposed by A.~Vikhlinin can reduce the particle background.  On
applying the filter and checking the image of events flagged by it, we
discovered that some source photons were being identified as
background particles. Therefore, we decided not to use such screening. 
We also found that cosmic ray afterglows add a negligible fraction to
the total event rate, and therefore did not remove the afterglow
correction included in the pipeline level~1 event file. 

The aspect offset for the A1991 observation is $< 0.1$~arcsec.
For source detection, we used the wavelet-based CIAO ``Wavdetect''
program with the source detection threshold set to $10^{-6}$.  
For all analyses of the diffuse cluster gas,
we used an image from which we eliminated the point sources so detected. 
We converted the counts image to a flux image by dividing it by an 
exposure map created using the aspect histogram and an instrument map. 

For spectral analysis, we accounted for the position--dependent
variations in scale and resolution of the energy of the recorded
events by appropriate choice of the Redistribution Matrix File (RMF)
and the Auxiliary Response File (ARF) for the source. We used the
``Acisabs'' program supplied by the Chandra X--ray Center to correct
for the degradation in the quantum efficiency of the ACIS detector at
low energies before performing spectral fits. The correction generates
a small but noticeable effect on temperature fits at low photon
energies, and affects determination of the equivalent hydrogen column
in the spectral models. 

For background subtraction or modeling, we used the source--free
extragalactic sky event file of M.~Markevitch\footnote{See
http://cxc.harvard.edu/contrib/maxim/acisbg/} to avoid issues with
spatial nonuniformity that might arise from using source--free sky
regions in the cluster observation itself.  We reprojected the
appropriate ACIS-S3 background file to match our observation. 

%
% ------------------------------------------------------------
%

\section{Morphological Analysis \label{sec:morph}}

\subsection{The {\it Chandra} Image}

\begin{figure}
\epsscale{1.18}
\plottwo{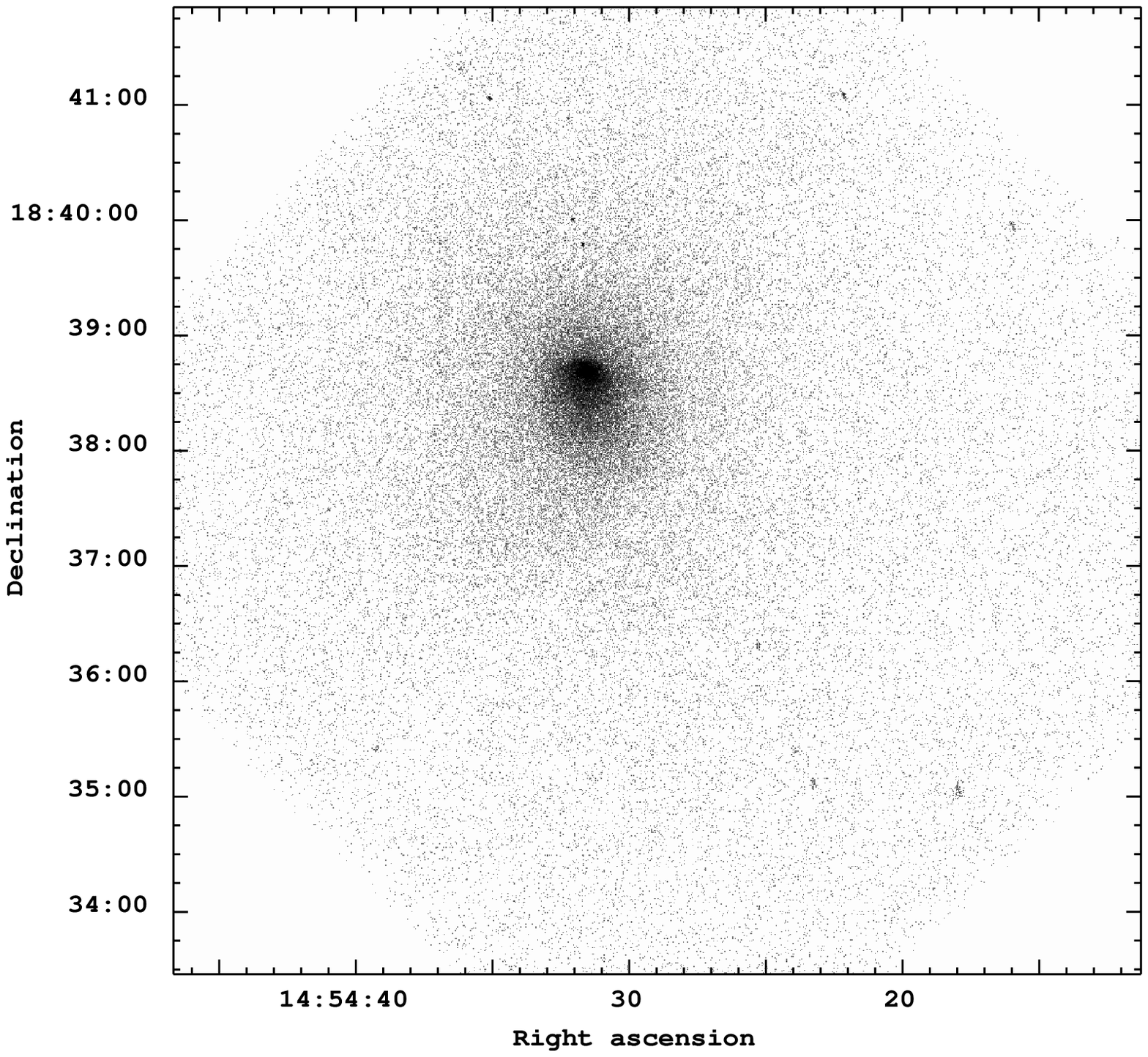}{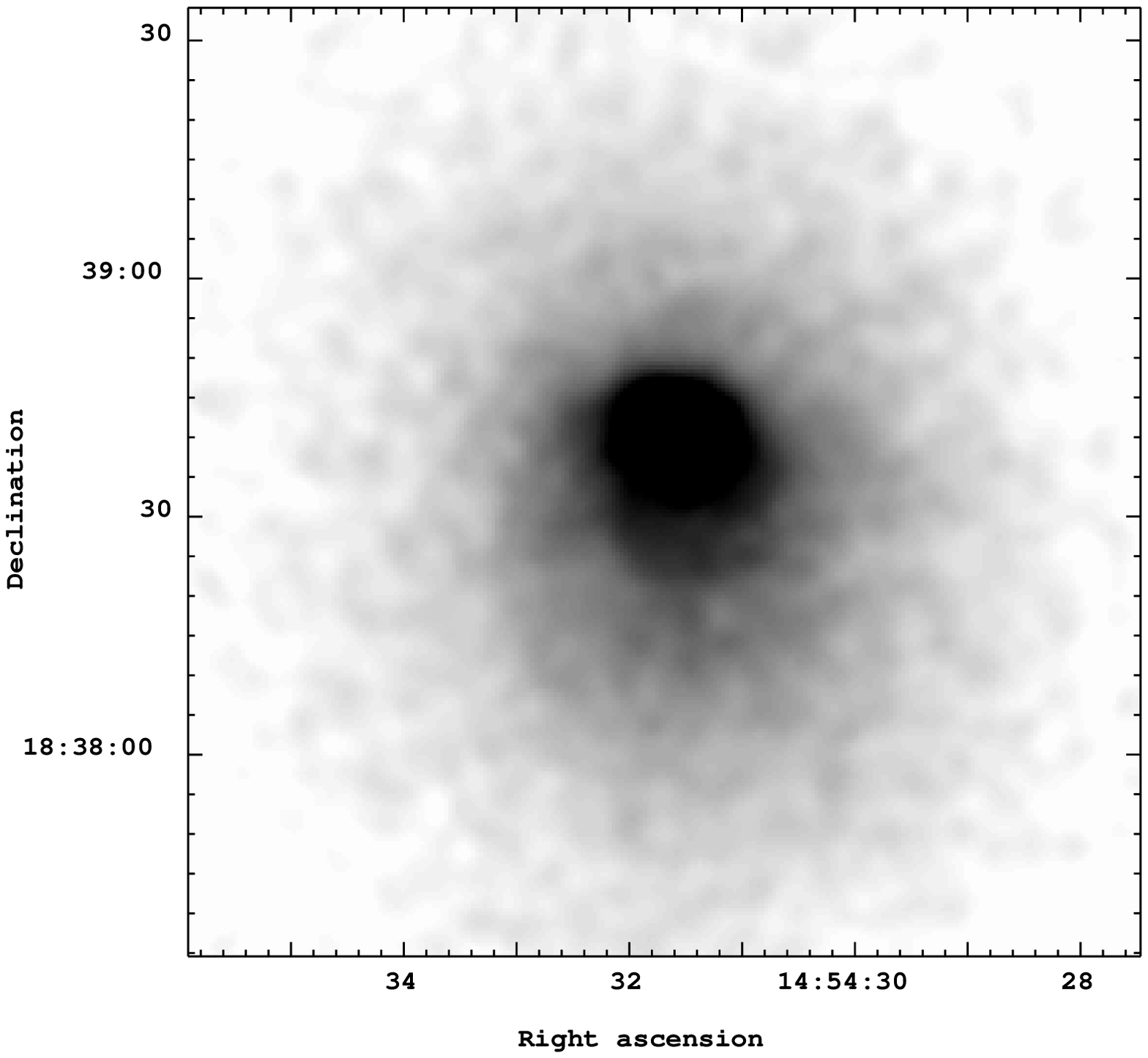}
\caption{{\it Top panel}: {\it Chandra} ACIS-S3 image of Abell~1991 
in the $0.3-10.0$~keV energy band.  
The raw image, with $0.49$~arcsec pixels, is $8$--arcmin on a side, 
corresponding to a linear size of $\sim 500$~kpc at the redshift 
of the cluster. {\it Bottom}: The central $2$~arcmin region of the same image, 
Gaussian smoothed to show the cometary structure.
The co-ordinates are J2000. 
\label{fig:clusimage1}}
\end{figure}

Figure~\ref{fig:clusimage1} (top panel) depicts the full--resolution 
{\it Chandra} ACIS-S3 image of A1991 in the $0.3-10.0$~keV band. 
Within the central area of $r \sim 150$~arcsec, the net count rate is 
$\approx 1.9\ {\rm count\ s}^{-1}$.  The X--ray emission of the cluster 
appears to be regular, except at the center. 
The X--ray centroid (J2000 RA, Dec=14:54:31.7, +18:38:35) and 
peak (J2000 RA, Dec=14:54:31.5, +18:38:42), determined over a roughly 
$100$~arcsec central region of the X--ray image, are offset by a few arcsec 
from the optical/radio position (J2000 RA, Dec=14:54:31.5, +18:38:32.0) 
of the brightest cluster galaxy (BCG), NGC 5778. 
At about $10$~arcsec north of the X--ray centroid, 
there is enhanced emission (shown in the lower panel of 
Fig.~\ref{fig:clusimage1} and discussed in \S~\ref{sec:knotmorph} below). 
Further, there appear to be weak ``edges'' in the surface brightness 
similar to those in several rich clusters studied with {\it Chandra} 
(e.g., Abell 3667; Vikhlinin, Markevitch, \& Murray 2002). 

Figure~\ref{fig:dss} shows the optical image from 
the Digitized Sky Survey overplotted with the contours of 
the {\it Chandra} X--ray image in the energy range $0.6-7.0$~keV. 
In addition to the diffuse ICM, discrete sources appear on 
the {\it Chandra} image; 17 are detected by the CIAO 
``Wavdetect'' program over the area of the ACIS-S3 CCD. 
We overplot these as diamonds in Fig.~\ref{fig:dss}.
Appendix A provides some details of the sources. 

\begin{figure}
\epsscale{0.8}\plotone{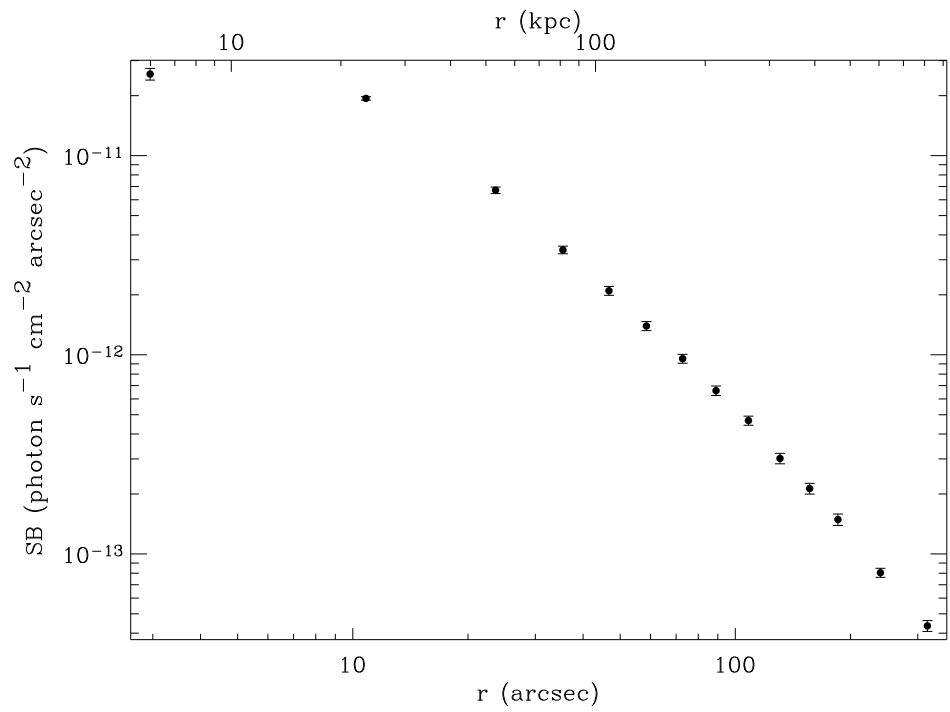}
\caption{Surface brightness over $0.6-7.0$~keV as a function of
distance from the optical center of the brightest cluster galaxy. The
error bars are at the $1 \sigma$ level. 
\label{fig:sb}}
\end{figure}

Figure~\ref{fig:sb} plots the cluster X--ray surface brightness
profile centered on the BCG, corrected for exposure and background,
and cleaned of point sources. In the image, the diffuse cluster
emission extends at least $180$~arcsec from the central galaxy.
White et al. (1997) found
a core radius of $r_c=200$~kpc for A1991 (corresponding to an angular
size of approximately $120$~arcsec using their cosmological parameters
of $H_0=50$ km s$^{-1}$ Mpc$^{-1}$ and $q_0=0.5$). In this work,
we study a similar $200$~kpc inner region of A1991. 

%
% -----------------------------
%
\subsection{Substructure in the ICM \label{sec:knotmorph}}

The full--resolution image of A1991 shows a complex structure
roughly $10$~arcsec north of the optical center of the BCG.  Such
knotty features have been frequently reported from {\it Chandra}
studies of cores of galaxy clusters, e.g., 
in 2A0335$+$096 (Mazzotta et al.~2003) and others. 

\begin{figure}
\epsscale{1.35}\plottwo{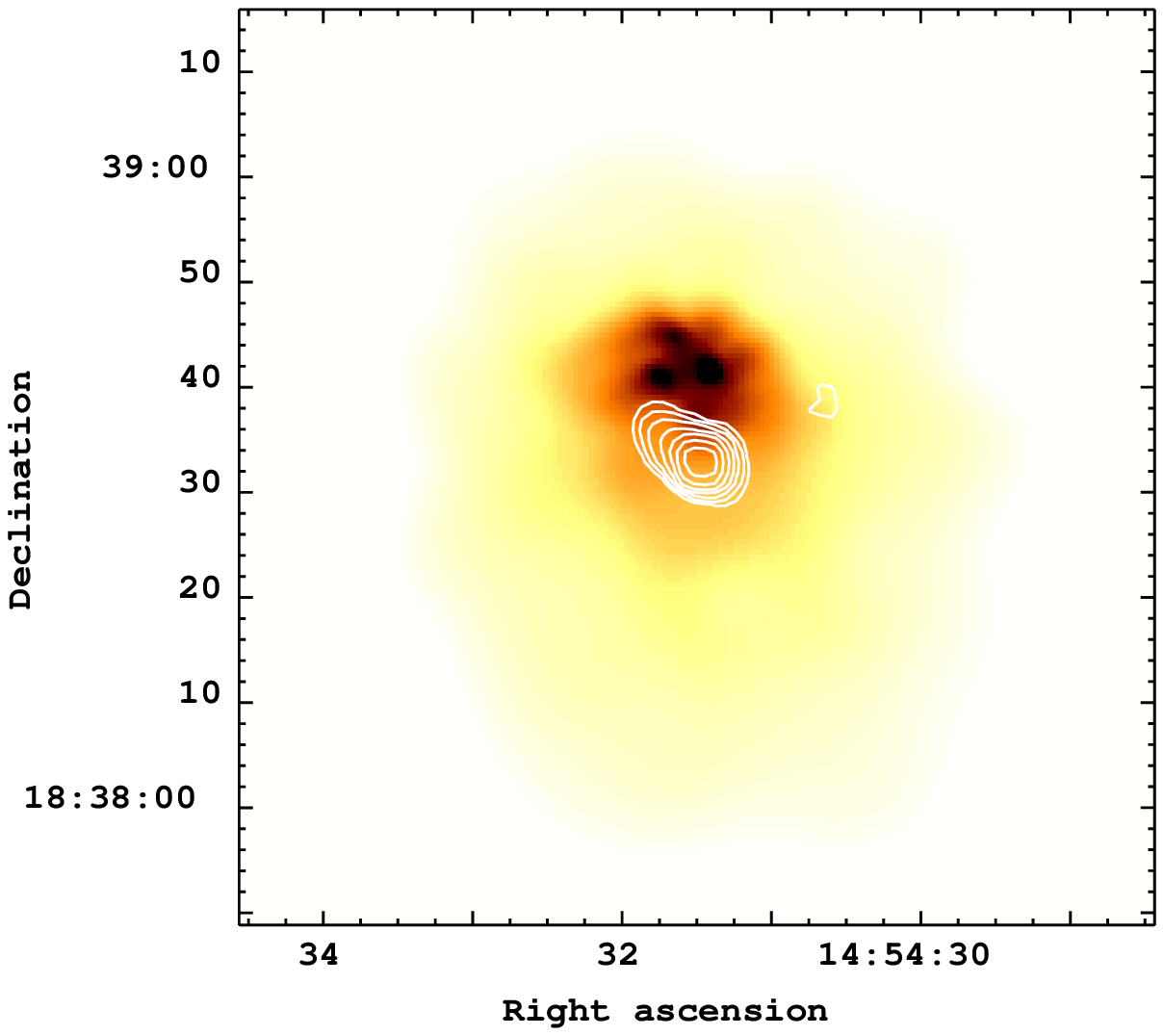}{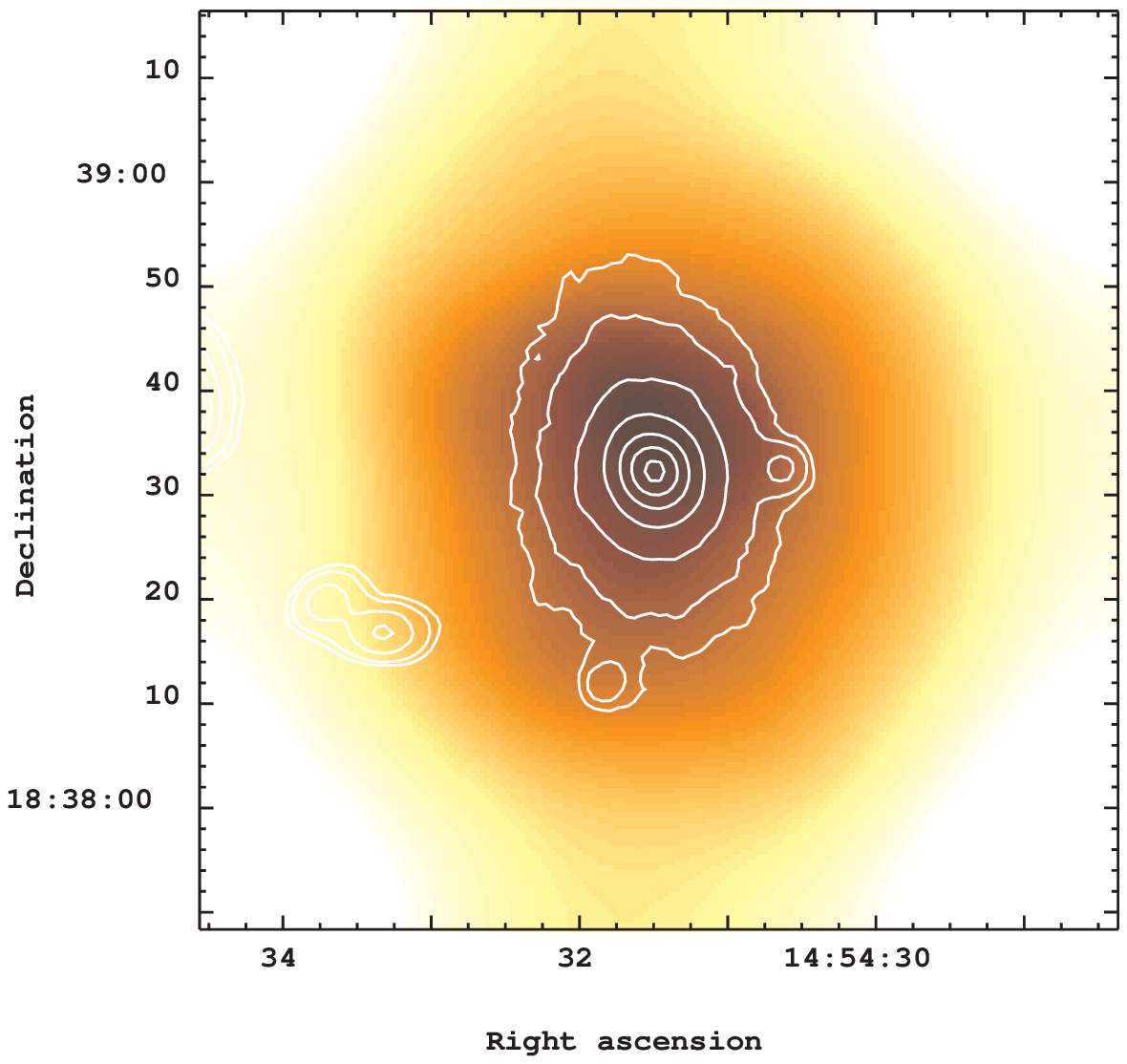}
\caption{Adaptively--smoothed {\it Chandra} images of the central
$\approx 1.5\times 1.5$~arcmin region of A1991, with $0.49$~arcsec pixels.
The top and lower panels are images in the soft ($0.3-2.0$~keV) and
hard ($2.0-7.0$~keV) X--ray bands, respectively, with arbitrary
scaling. The soft X--ray image has contours of the $20$-cm VLA radio
image overplotted on it. 
The X--ray knots lie adjacent to the radio contours at flux
levels of $\approx 1$~mJy per beam. The hard X--ray image is
overplotted with contours of the optical $B$ band image of the cluster
center. 
\label{fig:knot_soft_hard}}
\end{figure}

To study any possible temperature dependence of the structure, we
filtered the {\it Chandra} image by energy and created separate soft
($0.3-2.0$~keV) and hard ($2.0-10.0$~keV) band images. 
Figure~\ref{fig:knot_soft_hard} depicts the adaptively smoothed soft
and hard X--ray images of the $\approx 1.5$~arcmin square region
centered on the BCG. The centroids of the soft and hard emission are
within 1~arcsec of each other and of the centroid of the total
X--ray emission.  The hard emission has a nearly smooth, rhomboid
shape, while the $0.3-2.0$~keV map shows interesting structure in the
form of a region of enhanced soft X--ray emission located at the north
edge of the cD galaxy.  Indeed, this knotty structure shows little
corresponding emission above 2~keV; in \S~\ref{sec:knotspec}, we
demonstrate that it has a spectral identity distinct from that of the
cluster.  The structure (called the knots, henceforth) 
is slightly elliptical in projection, with
semi-major and semi-minor axes of approximately $6.7$~arcsec (7.7~kpc) and
$5.9$~arcsec (6.8~kpc), respectively.  It actually consists of
several finer structures on arcsecond scales that appear to be
connected. 
As the bottom panel of Fig.~\ref{fig:clusimage1} shows, the bright structure 
has a sharp edge to the north, and a less sharp one 
to the south. 
In Fig.~\ref{fig:knot_soft_hard}, we overplot contours of a
$21-$cm radio image acquired at the VLA on the soft X--ray image.
Interestingly, the X--ray knots lie adjacent to the radio source,
but the radio and X-ray emission appear to be unrelated to
each other.

%
% ------------------------------
%

\section{Spectral Analysis \label{sec:specan}}

We extracted spectra in pulse invariant
(PI) channels from the events file cleaned of point sources. We
grouped the spectra to have a minimum number of 20 counts per bin. We
generated background spectra from corresponding regions of the blank
sky background of Markevitch, except where indicated otherwise
(Sec. \ref{sec:knotspec}). For spectral model fitting, we
used the XSPEC program (Arnaud 1996).  Despite recent improvements,
the ACIS response below $0.5$~keV and around energies of $1.4-2.2$~keV
has calibration uncertainties.  To assess their significance, we
modeled the average spectrum both including and excluding this energy
interval.  Excluding the $1.4-2.2$~keV energies resulted in the
$\chi^2$ values becoming marginally better, but the best--fit
parameter values and 90\% confidence intervals remained virtually
unchanged from the fits over the full energy range.  In this work, we
report only results from spectral analyses over the full energy range
with the lower cut--off at $0.6$~keV and high--energy cut--off at
$7.0$~keV, or $4.0$~keV for the knots.

The soft X--ray spectrum is subject to absorption by neutral hydrogen
in the Galaxy.  We parametrize the interstellar X--ray absorption
using the XSPEC/WABS model that uses the absorption cross-sections of
Morrison \& McCammon (1983). We calibrate the abundance measurements
relative to the solar photospheric values of Anders \& Grevesse
(1989). 

We first study the average {\it Chandra} spectrum of the ICM in A1991
for comparison with the literature. Next, we profile the radial
variations in temperature, metallicity, pressure, electron
density. Subsequently, we study the spectral nature of the
knots (Sec. \ref{sec:knotspec}). 

%
% ------------------------------
%

\subsection{Average Cluster Spectrum \label{sec:avgspec}}

In order to compare the global properties of the cluster to
earlier, low resolution X-ray observations, we 
extracted a spectrum over the energy range
$0.6-7.0$~keV within the largest circular area fully circumscribed by
the ACIS-S3 chip without spilling over the CCD edges. This region,
centered on the optical position of the BCG, has a radius of about
190~arcsec ($\approx 215$~kpc), and contains about $70,000$
net counts.  We wish to point out at the outset that
due largely to the temperature decline in the
inner 100~kpc of the cluster, none of the model fits to the
global spectrum provide a satisfactory fit to the data.

We fitted the following models to the average spectral energy distribution: 
(i) MEKAL (Mewe, Gronenschild, \& van den Oord 1985; Liedahl,
Osterheld, \& Goldstein 1995), an absorbed single-temperature,
optically thin plasma model that includes line emissions from several
elements, and
(ii) MEKAL$+$MEKAL, a combination of two MEKAL models with
metallicities set to be identical.
For both the above models, we left the  temperature(s), metal
abundance and normalization free to vary but fixed the redshift. 

We ran the fitting programs twice, with the equivalent hydrogen column
density first left as a free parameter and next, frozen to the
Galactic value of $N_H = 2.25 \times 10^{20}\ {\rm cm}^{-2}$ 
(Bonamente et al.~2002).  
We found, however, that allowing $N_H$ to vary resulted in its being
poorly--constrained in most of the model fits, with fitted values 
sometimes significantly below Galactic.
While the presence of the lower values could be a reflection of the
models being unable to fully describe the average spectrum, it could also
be due to uncertainty in the correction to the ACIS quantum efficiency
at low energies. We repeated the fits for a slightly higher lower
energy cut--off of $0.8$~keV. The resulting values of the absorption column
density were again close to or below Galactic, and remained weakly constrained.
Below, we discuss only results after fixing the equivalent hydrogen absorption 
to the Galactic value in the WABS model. The errors we quote correspond to 
90\% confidence limits for one interesting parameter.

% ------------------------------

For the {\it Chandra} spectrum between $0.6-7.0$~keV, a
single--temperature MEKAL model fit gives a temperature of 
$kT = 3.23 \pm 0.09$~keV and chemical abundance of
$0.73 \pm 0.08$ solar. 
However, this model is not a good fit to the data with
$\chi^2=777.0$ for 355 degrees of freedom.
As we show in \S~\ref{sec:temp_prof}, there is a 
temperature gradient over this aperture, implying the ICM cannot 
be described as isothermal. In comparison, Ebeling et al.~(1996) 
estimated $kT=5.4$~keV over the {\it ROSAT} energy range of 
$0.1-2.4$~keV within a $5.5$~arcmin radius. 

The model that seems to best fit the average cluster spectrum (still
with $\chi^2_{\rm red}=1.443$ for 353 degrees of freedom) 
needs a combination of two MEKAL components. The best--fit temperatures 
for the cool and hot thermal components are $1.51 \pm 0.09$~keV 
and $7.87_{-0.86}^{+1.06}$~keV, respectively. The temperature for 
the cooler MEKAL component is comparable to that derived from 
deprojection analysis of the total spectrum by Stewart et al.~(1984) 
who found $kT=1.64$~keV over the {\it Einstein} $0.5-3.0$~keV band. 
White et al.~(1997) estimated $kT=5.3^{+0.4}_{-4.0}$~keV for 
the deprojected {\it Einstein} spectrum between $0.6-4.5$~keV 
over a $5$~arcmin radius. 
The chemical abundance, assumed to be identical for both
the MEKAL components, is $0.77 \pm 0.13$ solar,
similar to that determined by the single-temperature MEKAL fit. 
For this two--MEKAL model, the total unabsorbed flux
($0.6-7.0$~keV) is $1.19 \times 10^{-11}$\ erg~\ cm$^{-2}$~\ s$^{-1}$, 
with a little less than one half contributed by the soft component. 
This model computes temperatures and abundance that are 
virtually unchanged regardless of whether $N_H$ is fixed or free to vary.

%
% ------------------------------
%
\subsection{Radial Profiles of Spectral Parameters}

To estimate the radial variations in temperature, metallicity,
pressure, and electron density, we extracted X--ray spectra in
concentric, circular annuli starting from the optical center of
the BCG. Since the overall cluster emission is nearly circular, and as
we assume spherical symmetry in deprojection, circular annuli are a
reasonable choice. 
As we show in \S~\ref{sec:knotspec}, the soft X--ray knots in the cluster 
core have a spectrum that is different from their surroundings.  
So, in estimating the radial dependence of the spectral
parameters, we excised the knotty region from annuli inside 
about $20$~arcsec. In the deprojection (see below), we accounted for 
the volume missing due to these excised regions. 
Each annulus includes approximately $5000$ counts before background
subtraction. 

We modeled the projected spectrum between $0.6-7.0$~keV at each 
radial distance with an absorbed single--temperature MEKAL model. We
initially allowed the equivalent hydrogen column density, along with
the temperature, metallicity, and normalization to vary freely
in the fitting process. The best-fitting absorption value for many of
the annuli was, as in the case of the global spectrum (\S~\ref{sec:avgspec}),
close to or below the Galactic value. It is unlikely that there is
an actual decrement in the Galactic $N_H$ along the line of sight of A1991.
To avoid complications in interpreting further
results, we performed spectral fits with the
value of absorption frozen at the Galactic value. 
In projection, the inner annuli to approximately 15~arcsec of
the cluster center are not very well fitted with a single--temperature
MEKAL model (average reduced $\chi^2$ of $1.2$ for about 90 degrees of
freedom). However, the simple MEKAL model does represent the spectra
reasonably well at larger radii. 

We also performed deprojection of the spectra to account for
contributions from the outer regions of the cluster to its inner
regions. The deprojection algorithm first fits an absorbed
single-temperature MEKAL model to the spectrum of the outermost
annulus.  It then works inwards, fitting to each annulus a MEKAL model
in addition to a weighted combination of the best--fit models of all
the annuli exterior to it. The weights given to the spectra of the
outer annuli are commensurate with the volume contributions of the
total outer annuli to the current one. 
Note that we make no allowance for the fact that the outermost annulus
also has contribution from cluster emission outside it; therefore,
the fluctuations in the deprojected quantities for the outer one or 
two rings are not to be considered physical. We also found that the spectrum 
of the innermost approximately $3$~arcsec of the cluster was unstable
in deprojection due to the displacement of the centroid and the
peak of the X-ray emission from the chosen center. We therefore present 
deprojected spectral parameters outside this radius only.  The deprojected fits 
up to the outer two annuli have reduced chi-squared values
close to one.

%
% ------------------------------
%

\subsubsection{Temperature and Abundance \label{sec:temp_prof}}

\begin{figure}
\epsscale{1.00}\plotone{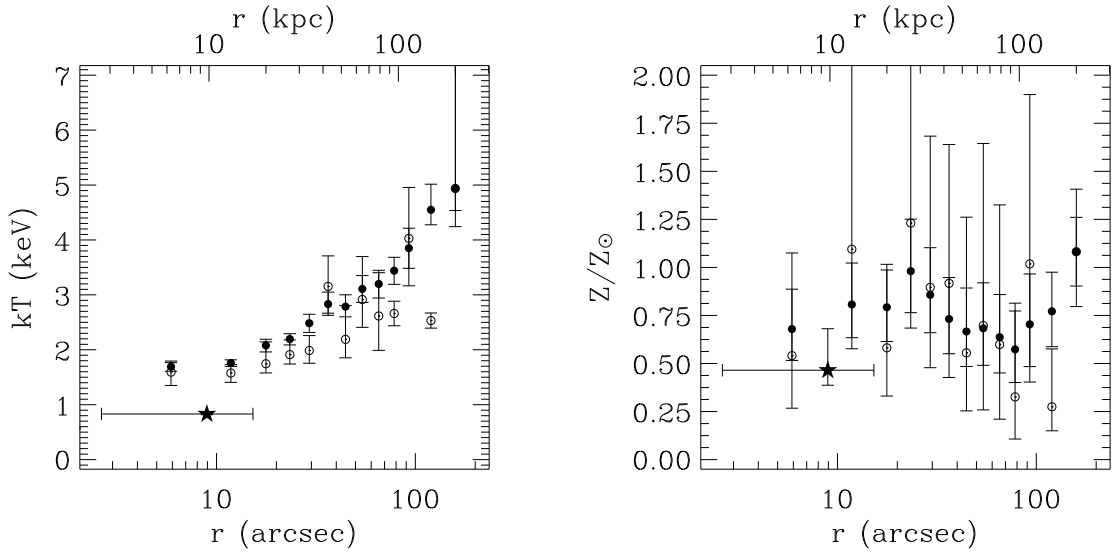}
\caption{Radial profiles of the gas temperature ({\it left panel}) and
metallicity ({\it right panel}) in A1991 obtained from fitting a
single-temperature MEKAL model to each annulus. Filled circles are for
the observed (projected) spectrum and open circles are from spectral
deprojection assuming spherical symmetry. Error bars show the $90\%$
confidence levels. The filled star at $r=9$~arcsec in each plot
denotes the corresponding best--fit parameter and its $1\sigma$
deviation for the knotty substructure studied in Sec.~\ref{sec:knotspec}. 
\label{fig:temp_abun_prof}}
\end{figure}

The left panel of Fig.~\ref{fig:temp_abun_prof} presents the radial
variation of the projected (filled circles) and deprojected (open
circles) ICM temperatures. Error bars show the $90\%$ confidence
intervals.  
The emission weighted temperatures of the various annuli bracket almost
a $3$~keV range over the cluster, with a slow positive radial gradient. 
The projected temperature drops from 
$kT \approx 3.9$ keV at a radial distance of about $95$~arcsec
($\approx 110$~kpc) to $kT \approx 1.7$~keV near the cluster center.  
After deprojection, which accounts for the overlying hotter layers, 
the temperature at all radii but the outermost (for which no correction
is applied) expectedly falls to a value lower than seen in projection.  
The three dimensional temperature falls to a low of $1.6$~keV in the center. 
The constraints on deprojected temperature, where the errors are correlated
between annuli, are not so strong as in the projected case.  

The right panel of Fig. \ref{fig:temp_abun_prof} shows  
the radial profile of the chemical abundance of the ICM.
Both the projected (filled circles) and deprojected (open circles)
profiles are noisy.
The metallicity values in the various annuli average around
$0.8\ Z_{\odot}$. 
This compares reasonably well with the overall cluster
abundance found from the best--fitting two--MEKAL model of the total 
cluster spectrum (\S~\ref{sec:avgspec}).  

% 
% ------------------------------
%
\subsubsection{Gas Density, Pressure and Entropy \label{sec:nepr_prof}}

\begin{figure}
\epsscale{0.90}\plotone{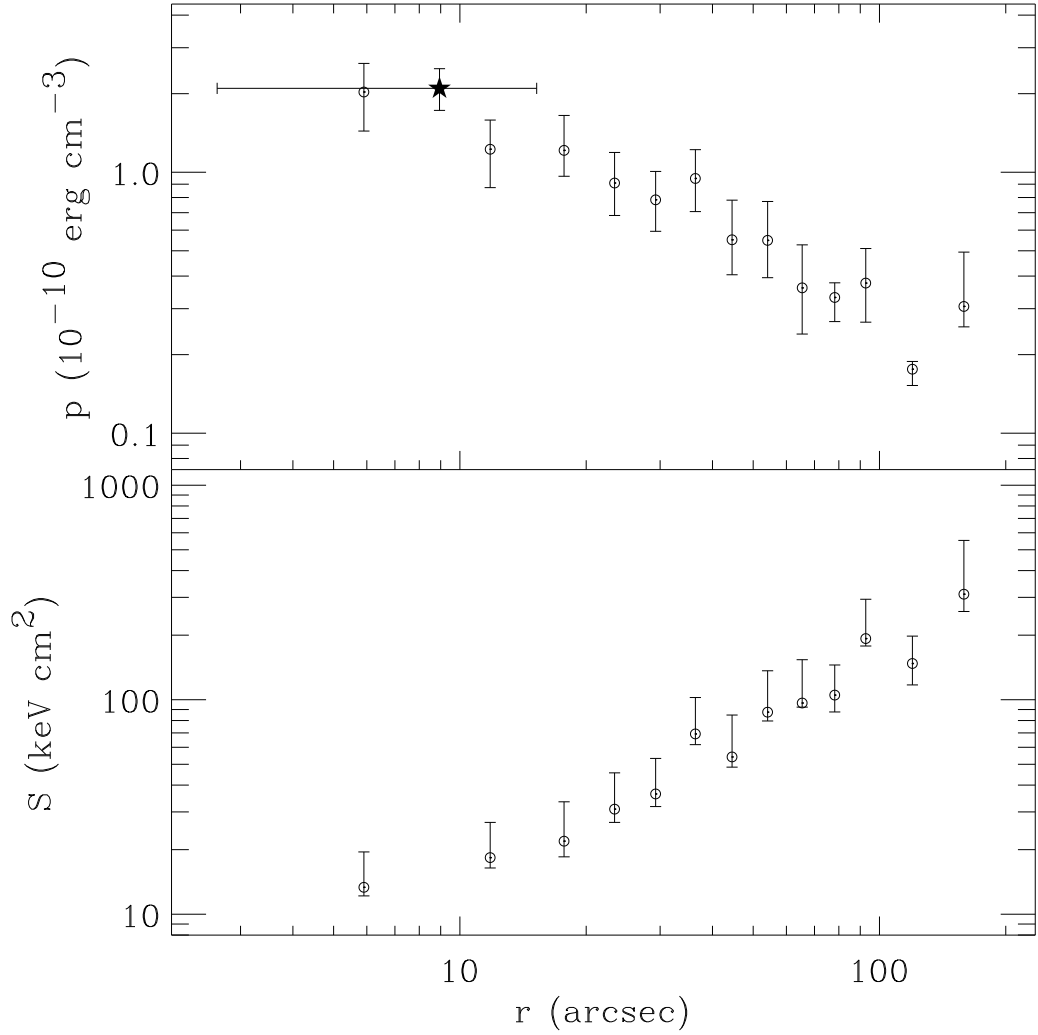}
\caption{Radial profiles of the gas pressure ({\it top panel}) and
entropy ({\it lower panel}) inferred from spectral deprojection. 
The filled star in the plot of pressure at $r=9$~arcsec represents the knots. 
\label{fig:prent}}
\end{figure}

From the deprojected spectral fits presented in
\S~\ref{sec:temp_prof}, we also compute the electron density, pressure,
and cooling time of the gas. The top panel of
Fig.~\ref{fig:netcoolmdot} plots the radial profile of the
deprojected gas density calculated from the MEKAL model
normalizations.  Over the inner $100$~kpc cluster region, 
where the deprojected gas temperature declines inwards, 
the gas density climbs by about a factor of ten, 
reaching a central value of $n_e \approx 0.04\ {\rm cm}^{-3}$. 
Further, there is no obvious constant density core in the cluster.  
Over the same annuli, the specific entropy (Fig.~\ref{fig:prent}, bottom
panel) of the ICM increases outward, as is seen in other cooling flow
clusters (e.g. David et al. 2001).
As the top panel of Fig.~\ref{fig:prent} shows, the ICM pressure climbs 
by nearly an order of magnitude from the exterior regions towards the cluster
center.  It approaches $2 \times 10^{-10}\ {\rm erg\ cm}^{-3}$ at the
location of the cD galaxy. This pressure is smaller by a factor of
several compared with clusters hosting massive cooling flows and
powerful radio sources like Hydra A (McNamara et al. 2000).

%
% ------------------------------
%
\subsubsection{Cooling Mass Deposition Rate \label{sec:maxcf}}

\begin{figure}
\epsscale{0.75}
\plotone{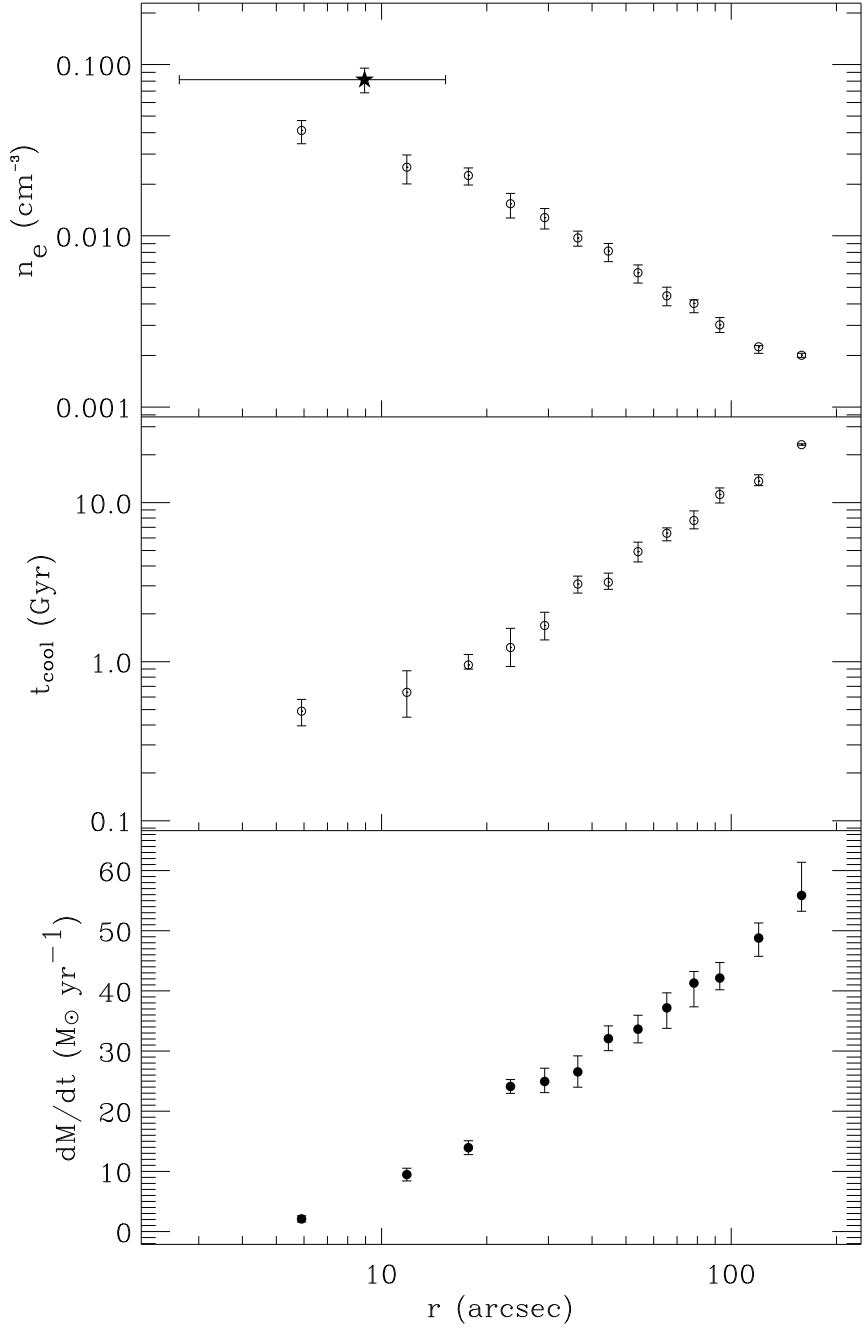}
\caption{Radial profiles of the deprojected gas density ({\it top panel}),
radiative cooling time ({\it middle}) and projected maximum mass deposition 
rate ({\it bottom panel}) in the ICM of A1991.  
The filled star at $r=9$~arcsec in the top plot denotes the corresponding
best--fit parameter and its $1\sigma$ deviation for the knots studied
in \S~\ref{sec:knotspec}. 
The maximum mass deposition rate assumes that gas is cooling to $0.01$~keV
in the cooling flow model (see \S~\ref{sec:maxcf}). 
\label{fig:netcoolmdot}}
\end{figure}

The middle panel of Fig.~\ref{fig:netcoolmdot} shows the radial
variation of radiative cooling time calculated from
the deprojected electron density and including the metallicity
dependence of the cooling function.  The central cooling time
is $\approx 500$ Myr, and rises to 
$\simeq 1$ Gyr at a radius of $\simeq 20$~kpc. 
A1991 is similar in this respect to other cooling 
flow clusters such as Perseus (Fabian et al.~2000) and 
Hydra A (McNamara et al.~2000; David et al. 2001; Nulsen et al. 2002). 

The short cooling times at the center of A1991 suggest that cooling to
low temperatures may be occuring in the core of the cluster.
However, high-resolution {\it XMM-Newton} spectra (e.g., Peterson et al. 2003; 
Kaastra et al. 2004) have shown that the bulk of the gas in cooling flow clusters 
cools from ambient temperatures down to only $\sim 2$~keV, with very little
cooling below 2~keV. 
The standard isobaric cooling flow model 
may thus be an over-simplification of the real cooling gas spectrum, 
and a cooling flow model with a minimum temperature cut-off provides a better 
description of the spectral data (see Molendi \& Pizzolato 2001).
Nevertheless, the {\it XMM-Newton} upper limits on the level of cooling below
X-ray temperatures are not restrictive, so substantial amounts of gas may yet
be cooling to low temperatures.
It is therefore reasonable to evaluate the possible maximum
mass deposition rates consistent with our low spectral resolution
data. 

We fitted the MKCFLOW model (Mushotzky \& Szymkowiak 1988) along with 
an isothermal (MEKAL) plasma component to the annular spectra,  
both modified by the WABS model with the line--of--sight absorption fixed 
to the Galactic value. 
The cooling flow model assumes that the gas cools isobarically from the ambient 
temperature (determined by the MEKAL component) to a low temperature 
(that we set to $0.01$~keV) at which it would no longer be detectable in X--ray. 
The cooling gas has the same abundance as the thermal component. 
Such a forced fit corresponds to a ``maximal cooling flow model'' as 
described by Wise et al.~(2004). 

We applied the statistical F-test to the chi-squared values of the
maximal cooling model versus the single temperature model for each of 
the annuli.
The cooling model provides a better fit to the data than a single 
temperature model within a $\sim 25$~arcsec radius.  
In this region, the cooling time is less than $1$~Gyr, and $\dot{M}$ is 
less than about $25$~\msunpa (a $1\sigma$ upper limit; see below).
At larger radii, single temperature models can adequately describe
the data, so the MKCFLOW component is not necessary.  
However, in several of the inner annuli where the maximal cooling
provides a statistically improved fit over the single MEKAL model,
a two--temperature (MEKAL$+$MEKAL) model (modified by WABS absorption) 
performs even better.
But it is not obvious what the physical interpretation of such a
two--temperature model would be. 

The lower panel of Fig.~\ref{fig:netcoolmdot} provides the results of 
the maximal cooling flow model fits. It depicts the radial profile 
of the projected cumulative mass deposition rate, $\dot{M}_{\rm max}$, 
within all the annuli inside a particular distance from the center. 
We find  $\dot{M}_{\rm max}$ rises from $\sim 1~$\msunpa to 
$\approx 51~$\msunpa over the central $120$~arcsec region. 
These values compare well with the estimates of
$\dot{M}=36^{+36}_{-11}$\msunpa of White~et~al.~(1997) 
and Stewart et al.~(1984) who found $\dot{M}=115$~\msunpa where the
cooling time is less than $5$~Gyr. 
We wish to emphasize that these mass deposition rates do not imply
that gas must be cooling at these rates. Rather, they represent
the maximum rates at which gas can be cooling to low temperatures 
for the model to remain consistent with the data.

The projected maximal mass deposition profile 
shows a break near $30$~arcsec from the center. 
Note, however, that no prominent corresponding break in the
X-ray brightness distribution is visible in Fig.~\ref{fig:sb}. 
Allen et al.~(2001) suggest that the age of cooling flows could be
identified with the cooling time of the ICM at the radius of such a
break in the profile of hardness ratio or mass deposition rates
derived from image deprojection analysis.  If we apply this method to
the break in the spectrally--determined cooling mass deposition rates,
we estimate an age of $\simeq 1$~Gyr for the cooling flow in A1991. 

%
% ------------------------------------------------------------
%

\subsection{Spectrum of the Cool Central Structure \label{sec:knotspec}}

The knotty region of enhanced X--ray emission described in
Sec~\ref{sec:knotmorph} seems largely confined to the soft X--ray band
(Fig.~\ref{fig:knot_soft_hard}) and, as we will show,
has a distinct spectral character. We extracted its X--ray spectrum
and fitted a thermal model to it in the energy range $0.6-4.0$~keV.
In order to subtract the cluster contamination from
the spectrum of the knots, we extracted 
spectra of the ICM from five regions at varying azimuthal angles
located along the circle
centered on the BCG with a radius terminating at the knots.
The knot spectrum has approximately $3000$ net counts after subtracting
the cluster background. 

\begin{figure}
\centering\includegraphics[width=10cm,angle=-90]{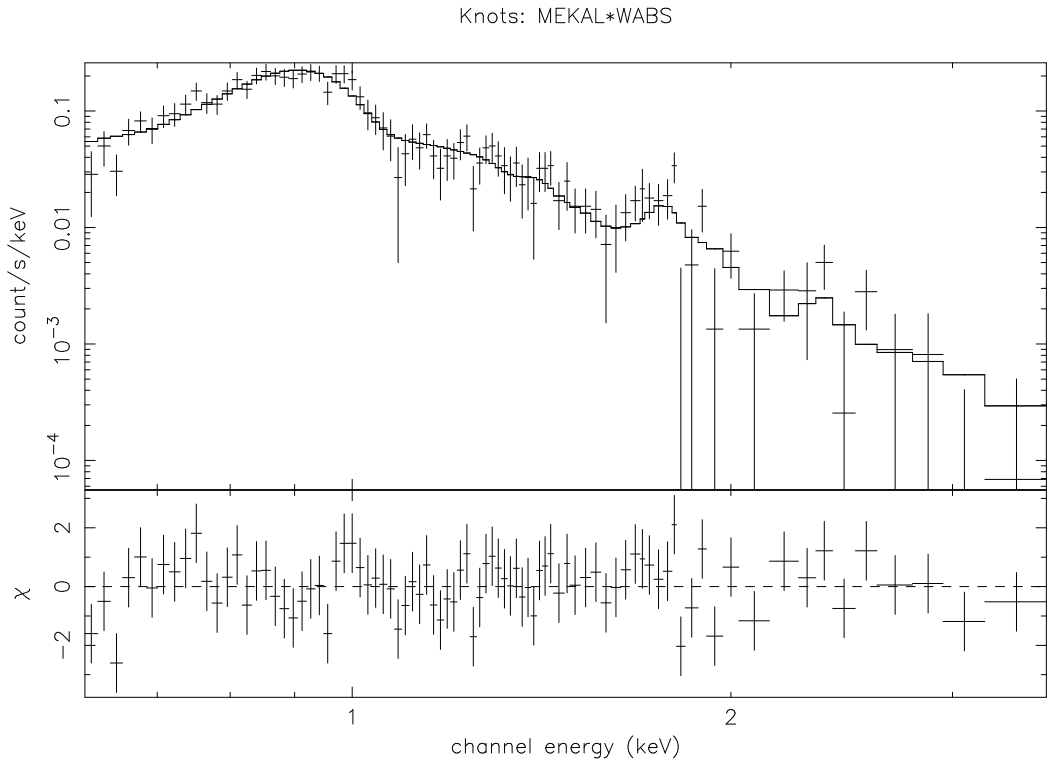}
\caption{Representative spectrum of the knotted substructure, 
overplotted with the best--fitting MEKAL model with
equivalent hydrogen column density fixed to the Galactic value. 
\label{fig:knotspec}}
\end{figure}

We fitted WABS absorbed single-temperature MEKAL models (with redshift
set to that of A1991) to the spectrum of the knots using in sequence
each of the five background regions.  Again, we adopted
the Galactic absorption.  
Figure~\ref{fig:knotspec} shows a representative spectrum of the knots
between $0.6-4.0$~keV overplotted with such an absorbed MEKAL model.
The fits are formally satisfactory in quality, with
$\chi^2/83$~d.o.f.\ $\approx1$.  
We quote the mean value of the five sets of the best--fit parameters 
and the 90\% confidence intervals.
We find that the substructure has a temperature of 
$0.83 \pm 0.02$~keV 
and metal abundance of $0.46_{-0.08}^{+0.22}\ Z_{\odot}$. 
We conclude that the knots are cooler than their surroundings
(see Fig.~\ref{fig:temp_abun_prof}), as also evidenced by the
comparison of the soft and hard X--ray images in
Fig.~\ref{fig:knot_soft_hard}. 
The total $0.6-4.0$~keV flux is 
$2.8 \pm 0.2 \times 10^{-13} \rm{erg~\ cm}^{-2}$~s$^{-1}$, 
where the error represents the $1\sigma$ deviation of the five measurements. 

We estimated the volume emission measure for the knots using 
the mean of the MEKAL model fit normalizations.  
From this and the mean knot temperature, we derive the gas density 
and pressure of the knots.  We compute the volume of the knots 
assuming the structure is an ellipsoid with the line--of--sight radius 
equal to the semi--minor axis of the ellipse. For semi--major and 
semi--minor axes of $6.67$ and $5.96$~arcsec, respectively, the knotty 
structure occupies $\approx 4.33\times 10^{67}\ {\rm cm}^3$. 
Then, the lower limit on the projected average
gas density within this volume, computed assuming the filling factor is
unity, is $n_e= 8.16_{-1.32}^{+1.37}\ \times 10^{-2}\ {\rm cm}^{-3}$.  
The mass of this substructure, assuming a mean particle mass of 
$\mu = 0.6$ times the proton mass, is
$3.4 \pm 0.6\ \times 10^9 M_{\sun}$, comparable to a small galaxy. 
The knot pressure is 
$p=2.10_{-0.37}^{+0.40}\ \times 10^{-10}\ {\rm erg\ cm}^{-3}$. 
The surrounding regions, which are about twice as hot (at $kT=1.58$~keV),
have a mean density $n_e=3.30_{-0.59}^{+0.52}\ \times10^{-2}\ {\rm cm}^{-3}$ 
and gas pressure $p=1.62 \pm 0.47\ \times 10^{-10}\ {\rm erg\ cm}^{-3}$ 
(as derived from interpolating the respective deprojected quantities from
Sec.~\ref{sec:nepr_prof}). Thus, the gas pressure in the cool knotty structure 
is $1.30 \pm 0.45$ times that of its surroundings. 
The primary source of error in our
determination of the gas properties of the knots is systematic:
the selection of the background region.
When our annular apertures are
centered on the X--ray centroid, rather than on the 
optical and radio co-ordinates of the cD galaxy, the background locations 
differ significantly
from the cluster center (i.e., to regions of lower pressure) on the
side opposite the knots.  In this instance, 
the ratio of the knot pressure to the surrounding pressure  
increases from $1.3$ to $1.9$.  This excess pressure 
implies a substantial peculiar velocity for the central structure,
as we discuss below.  

%
% --------------------------------------------------------
%
\subsection{Kinematics of the Cool Central Structure \label{sec:knotkine}}

The distinct character, high density, and cometary appearance of the cool, 
knotty structure suggest that it is moving northward through the core
of the cluster.  If so, we can apply the pressure ratio to 
estimate the peculiar speed of the structure.
If gas at the stagnation point in front of the moving knots had
the same properties as the gas at the same radius prior to being
disturbed by the knots, then, assuming a steady flow around the knots,
Bernoulli's theorem gives 
$H_0 + v_{\rm k}^2/2 = H$, 
where $H$ is the specific enthalpy of gas at the stagnation point, 
$H_0$ is the specific enthalpy of undisturbed gas at the same radius 
and $v_{\rm k}$ is the speed of the knots.  
If the knots are subsonic, gas is compressed adiabatically as it flows 
to the stagnation point, so that
$H/H_0 = (p/p_0)^{2/5}$, 
where $p$ is the pressure at the stagnation point and 
$p_0$ the pressure of the undisturbed gas.  
Also, $H_0 = 5kT_0 /(2 \mu m_{\rm H}) = 3s_0^2/2$, where 
$T_0$ is the temperature and $s_0$ the sound speed of the undisturbed gas, 
so that Bernoulli's theorem may be written $1 + m^2/3 = (p/p_0)^{2/5}$, 
where $m$ is the Mach number of the knots.  
If the pressure at the stagnation point equals the pressure in the knots, 
using the pressure ratio $p/p_0 = 1.30$ determined above, this gives 
the Mach number of the knots as $m=0.57$.  The Mach number corresponding 
to the maximum pressure ratio, $p/p_0 = 1.74$, is $m = 0.86$.  
The allowed range includes $p/p_0=1$, so that the knots could also be 
at rest in the ICM, although the appearance of the substructure
suggests otherwise.  We conclude, then, that the knots are almost
certainly moving  through the core of the cluster
at a substantial fraction of the sound speed.

The appearance of the cool structure suggests that it may be in the
final stages of breaking apart.  Pressure variations due to internal flow
are limited to the ram pressure of the external flow, i.e., 
the overpressure of the knots.  Since the knots are overpressured
by a factor of $1.3$,
the internal pressure variations must be modest, so that the density variations 
must be chiefly due to entropy variations within the structure.  
The ion mean-free-path in the ICM determined by Coulomb collisions, for 
$kT=1.6$ keV and $n_e = 0.033 \rm\ cm^{-3}$, is $\lambda \simeq 11$ pc, 
so that the Reynolds Number of the external flow is 
$R \simeq m r_{\rm k} / \lambda \sim 500$, where $r_{\rm k}$ is the radius
of the knotty structure, and
may be considerably larger if magnetic fields suppress ionic
viscosity.  Thus it is plausible that shear instability has driven
complex flow within the knotted substructure that is in the process 
of destroying it. 

Dynamical instabilities and the bulk motion of the core gas relative
to the rest of the cluster can also cause substructures such as edges
and blobs in the ICM (see e.g., Markevitch et al. 2000;  Mazzotta et
al. 2003). However, the central galaxy of A1991 has no significant
radial velocity offset with respect to the cluster mean (Beers et al.
1991), though there could be a component perpendicular to the line of
sight.  Assuming the thermal knots are presently part of the cluster but
originated as a distinct entity, the X--ray derived mass, gas
temperature, pressure and density of the cool structure are consistent 
with its being the remnant of a galaxy group that is falling into A1991.  
The group would be subjected to ram--pressure stripping as it traverses 
the cluster, but its dense (low--entropy) core would survive the passage. 
There is a hint of a front not far from the knots which may
represent the wake of the stripped gas. 

%
% ------------------------------------------------------------
%
\section{Conclusions \label{sec:concl}}

We have presented a new {\it Chandra} observation of
the A1991 galaxy cluster.
We find that the gas in the central 200~kpc 
has an average temperature of $1.51 \pm 0.08$~keV, 
and an average metallicity of $0.77\ Z_{\sun}$.
Like other cooling flow clusters, the temperature of
the gas in A1991 drops in the core.  The deprojected,
spatially--resolved gas temperature falls from 4~keV at 100~kpc
to 1.6~keV in the central 10~kpc. 
Spectral deprojection shows that the chemical abundances are roughly
constant across the central 100 kpc of the cluster.
The central cooling time 
is $t_{\rm cool} \approx 5\times 10^8$~yr, 
and rises to $\sim 10$~Gyr beyond $100$~kpc.
The mass deposition rate is about $25$~\msunpa 
within the central 20~arcsec of the cluster.  

We discovered a cool central structure defined by a collection
of bright X--ray knots projected $10$~kpc north of the nucleus
of the cD galaxy. The structure has a mass of $(3.4 \pm 0.6)\times
10^9\ M_{\sun}$, and is $\sim 1$~keV cooler than the surrounding gas.
It appears to be a small galaxy group that is breaking apart 
as it plunges northward through the core of the cluster at 
a substantial fraction of the sound speed.

%
% ------------------------------------------------------------
%

\acknowledgments

This research was supported by Chandra General Observer Award GO2-3161X
and NASA Long Term Space Astrophysics Grant NAG4-11025.
P. E. J. N. acknowledges support from NASA Grant NAS8-01130.
Support for E. L. B. was provided by NASA through the {\it Chandra}
Fellowship Program, grant award number PF1-20017, under NASA contract 
number NAS8-39073.

%
% ------------------------------------------------------------
%

\appendix

\section{Point Sources \label{sec:ptsrc}}

\begin{figure}
\epsscale{1.0}\plotone{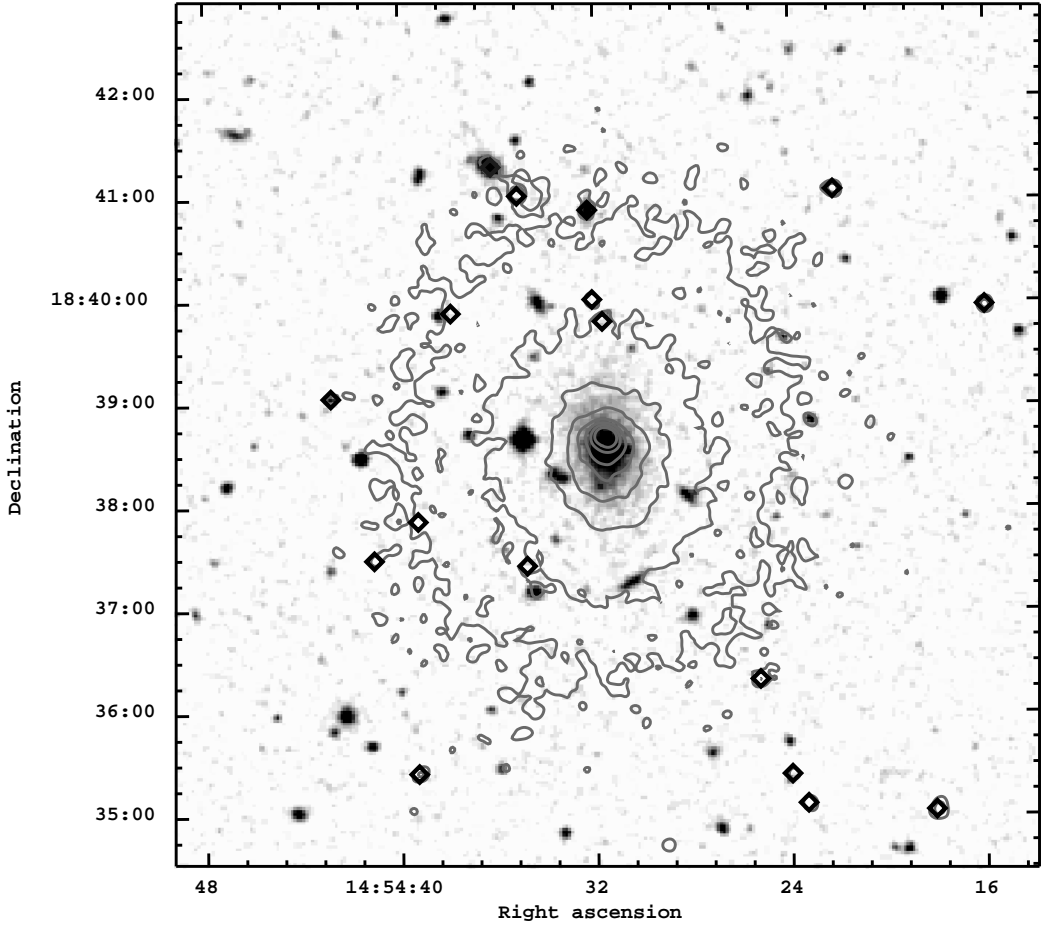}
\caption{Digitized Sky Survey optical image of the central 8~arcmin of
A1991 overplotted with the $0.6-7.0$~keV X--ray contours.
The co-ordinates are epoch J2000. Also overplotted as diamonds
are the discrete sources found by the Wavdetect program and listed
in Table~A.1.
\label{fig:dss}}
\end{figure}

Superposed on the cluster diffuse emission are seventeen discrete sources
that were detected using the CIAO ``Wavdetect'' program.
For source identification, we cross--correlated the X--ray co-ordinates
output from Wavdetect with the USNO B1.0 catalog positions of optical 
sources within approximately $1$~arcsec. 
Table~\ref{tab:ptsrc} presents the positions, count rates 
in the $0.3-10$~keV energy interval, and, when possible, the USNO
counterpart. 
In Fig.~\ref{fig:dss}, the diamond symbols overplotted on the optical
image represent the detected X--ray sources.

%
% ------------------------------------------------------------
%

%
% ------------------------------------------------------------
%

\clearpage

% -------------- Table of Knot parameters -------------

\begin{deluxetable}{l l}
\tablecolumns{2}
\tablewidth{0pt}
\tablecaption{Parameters of the knotted substructure. 
Values quoted are the mean and 90\% confidence intervals from five 
measurements 
using different backgrounds at roughly the same radial distance from 
the central galaxy as the knots. 
\label{tab:knot1}} 
% -------
\tablehead{
\colhead{Parameter}	& \colhead{Mean}	
} 
\startdata
\tableline
  & \\
Temperature 	 & $0.83 \pm 0.02$~keV \\
Electron density & $8.2 \pm 1.4\ \times 10^{-2}\ {\rm cm}^{-3}$ \\
Pressure 	 & $2.1 \pm 0.4\ \times 10^{-10}\ {\rm erg\ cm}^{-3}$ \\
Mass 		 & $3.4 \pm 0.6\ \times 10^9\ M_{\odot}$ \\
  & \\
\tableline 
\enddata
\end{deluxetable}

\clearpage
%
% -------------- Table of Point Sources --------------
%
\begin{deluxetable}{l l c c}
\tablecolumns{4}
\tablenum{A.1}
\tablewidth{0pc}
\tablecaption{Parameters of the point sources detected by the Wavdetect 
program within the area of the ACIS-S3 CCD. The count rate is calculated
over the $0.3-10.0$~keV energy range. Objects with USNO optical 
counterparts within 1~arcsec are noted. 
\label{tab:ptsrc}}
% -------
\tablehead{
\colhead{R.A. (2000)} & \colhead{Dec (2000)} 
	& \colhead{area}	& \colhead{count rate} \\
\colhead{hh:mm:ss.ss} & \colhead{dd:mm:ss.s} 
	& \colhead{arcsec$^2$}	& \colhead{10$^{-3}$\ ct s$^{-1}$}}
\startdata
\tableline
  & \\
14:54:16.01 & +18:39:58.1 & 18 & 0.9 \\
14:54:17.99 & +18:35:02.9 & 137 & 2.5 \\
14:54:22.15 & +18:41:05.1 & 25 & 1.4 \\
14:54:23.25 & +18:35:07.5 & 35 & 0.9 \\
14:54:23.94 & +18:35:23.5 & 34 & 0.5 \\
14:54:25.30 & +18:36:18.7 & 31 & 0.9 \\
14:54:31.71 & +18:39:47.7 & 23 & 1.7 \\
14:54:32.06 & +18:40:00.2 & 13 & 1.3 \\
14:54:32.22 & +18:40:53.4 & 22 & 0.7\tablenotemark{a} \\
14:54:34.66 & +18:37:26.2 & 30 & 1.1\tablenotemark{b} \\
14:54:35.10 & +18:41:03.5 &  9 & 2.6 \\
14:54:36.17 & +18:41:18.8 & 19 & 0.5\tablenotemark{c} \\
14:54:37.82 & +18:39:55.7 & 15 & 0.5 \\
14:54:39.14 & +18:37:54.4 & 19 & 0.6 \\
14:54:39.27 & +18:35:24.8 & 26 & 1.0 \\
14:54:41.01 & +18:37:29.1 & 18 & 0.4 \\
14:54:42.84 & +18:39:03.4 &  8 & 0.3\tablenotemark{d} \\
  & \\
\tableline
\enddata
\tablenotetext{a}{$m_B=16.1$~mag; stellar}
\tablenotetext{b}{$m_B=21.5$~mag; stellar}
\tablenotetext{c}{$m_B=15.1$~mag; extended}
\tablenotetext{d}{$m_B=19.1$~mag; non-stellar}
\end{deluxetable}
%

%
% ------------------------------------------------------------
%
\clearpage


% 
% ------------------------------------------------------------
% 
\end{document}